\input{aipcheck}

\documentclass[
    ,final            % use final for the camera ready runs
  ]
  {aipproc}

\layoutstyle{8x11single}

\begin{document}

\title{Neutron Skins and Neutron Stars}


\author{J. Piekarewicz}
 {address={Department of Physics, Florida State University,
                  Tallahassee, FL 32306-4350, USA}}

\begin{abstract}
 The neutron-skin thickness of heavy nuclei provides a fundamental link to 
 the equation of state of neutron-rich matter, and hence to the properties of 
 neutron stars. The Lead Radius Experiment (``PREX'') at Jefferson Laboratory 
 has recently provided the first model-independence evidence on the existence 
 of a neutron-rich skin in ${}^{208}$Pb. In this contribution we examine how 
 the increased accuracy in the determination of neutron skins expected 
 from the commissioning of intense polarized electron beams may impact 
 the physics of neutron stars.
\end{abstract}

\maketitle

%%%%%%%%%%%%%%%%%%%%%%%%%%%%%%%%%%%%%%%%%%%%
%% MAINMATTER
%%%%%%%%%%%%%%%%%%%%%%%%%%%%%%%%%%%%%%%%%%%%

\section{Neutron Skins}

The {\sl Lead Radius EXperiment} (``PREX'') at the Jefferson Laboratory has 
used parity-violating electron scattering to probe the weak-charge density of 
$^{208}$Pb\,\cite{Abrahamyan:2012gp,Horowitz:2012tj}. Given that the weak 
charge of the neutron is much larger than that of the proton, parity-violating 
electron scattering provides a clean probe of neutron densities that is free 
from strong-interaction uncertainties\,\cite{Donnelly:1989qs}. By invoking
some mild assumptions, PREX provided the first largely model-independent
determination of the neutron radius ($R_{n}$) of ${}^{208}$Pb. Since charge 
radii of stable nuclei are known with exquisite accuracy\,\cite{Angeli:2013}, 
PREX effectively provided the first clean evidence in favor of a neutron-rich 
skin in ${}^{208}$Pb\,\cite{Abrahamyan:2012gp,Horowitz:2012tj}:
%%%
\begin{equation}
r_{\rm skin}^{208}\!=\!R_{n}\!-\!R_{p}\!=\!{0.33}^{+0.16}_{-0.18}~{\rm fm}.
\end{equation}
%%%
Although PREX demonstrated excellent control of systematic errors,
unforeseen technical problems resulted in time losses that significantly 
compromised the statistical accuracy of the measurement. Fortunately,
the PREX collaboration has made a successful proposal for additional 
beam time, so that the original 1\% goal ($\!\pm0.06$\,fm) may be 
attained\,\cite{PREXII:2012}. Given that PREX demonstrated that 
model-independent measurements of weak-charge densities are now 
feasible, it is pertinent to ask whether a measurement in a different
neutron-rich nucleus could prove advantageous. Indeed, the case of
${}^{48}$Ca seems particularly attractive for several reasons. First, 
${}^{48}$Ca is
a doubly-magic nucleus that is already within the reach of ab-initio 
calculations\,\cite{Hagen2012b}. Thus, the {\sl Calcium Radius 
EXperiment} (``CREX'') could provide a critical bridge between ab-initio 
approaches and density-functional theory, that remains as the sole
theoretical alternative for the calculation of the properties of medium to 
heavy nuclei. Second, by providing this kind of bridge, CREX will help 
elucidate the character of the three-nucleon force, which plays a critical 
role in determining the limits of nuclear existence.  
Finally, CREX together with PREX-II will provide calibrated anchors 
for hadronic measurement of neutron skins at radioactive beam
facilities. In essence, PREX-II in combination with CREX will place
much-needed constrains on the poorly-known {\sl isovector} sector 
of the nuclear energy density functional, namely, that sector in which
protons and neutrons contribute with opposite sign. Note that a typical 
example of such a quantity is the isovector density 
$\rho_{{}_{\!1}}\!=\!\rho_{p}\!-\!\rho_{n}$ which---unlike the {\sl isoscalar} 
density $\rho_{{}_{0}}\!=\!\rho_{p}\!+\!\rho_{n}$---is fairly small for stable 
nuclei.

\begin{figure}[h]
 \includegraphics[height=7cm]{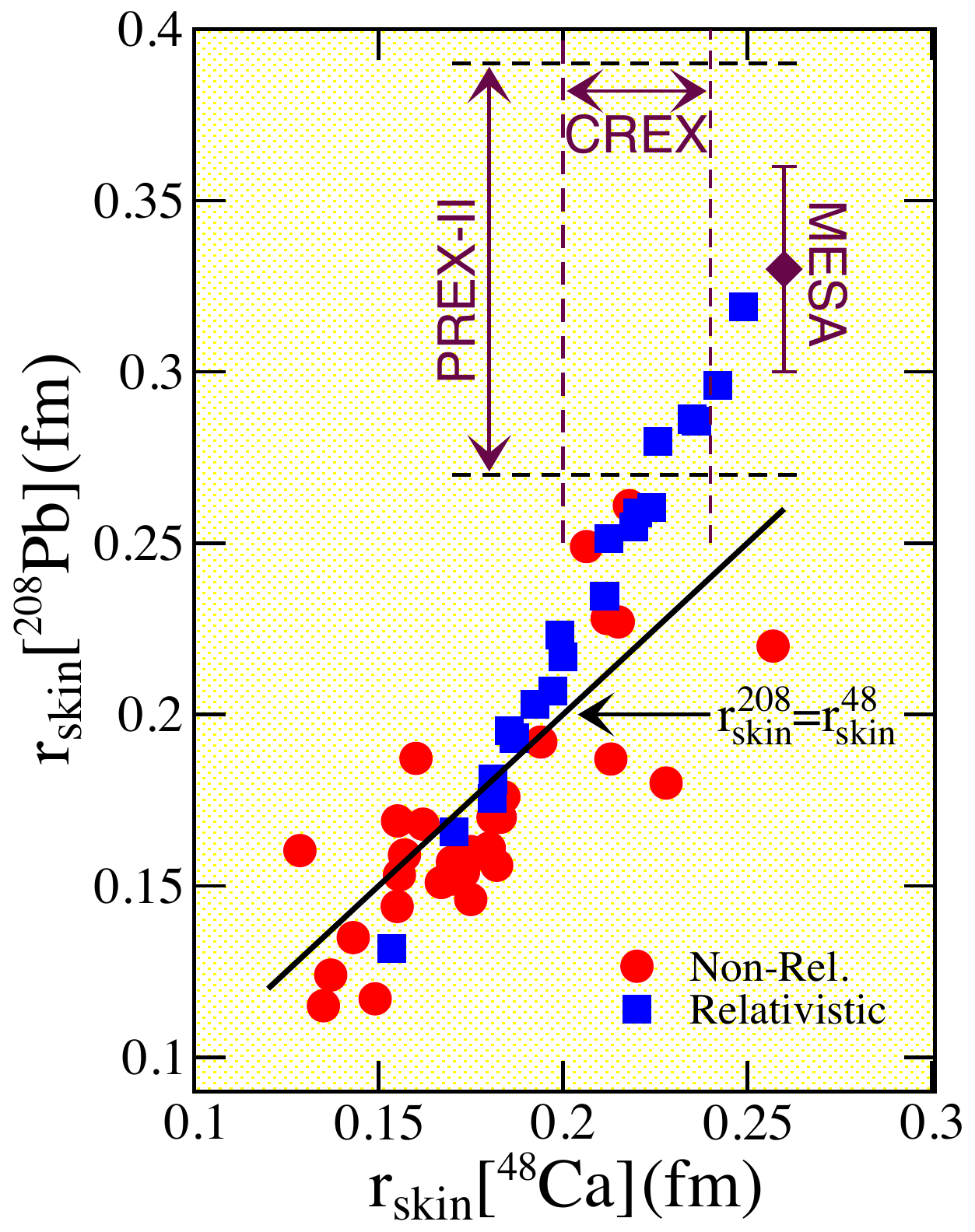}
  \hspace{0.75cm}
 \includegraphics[height=7cm]{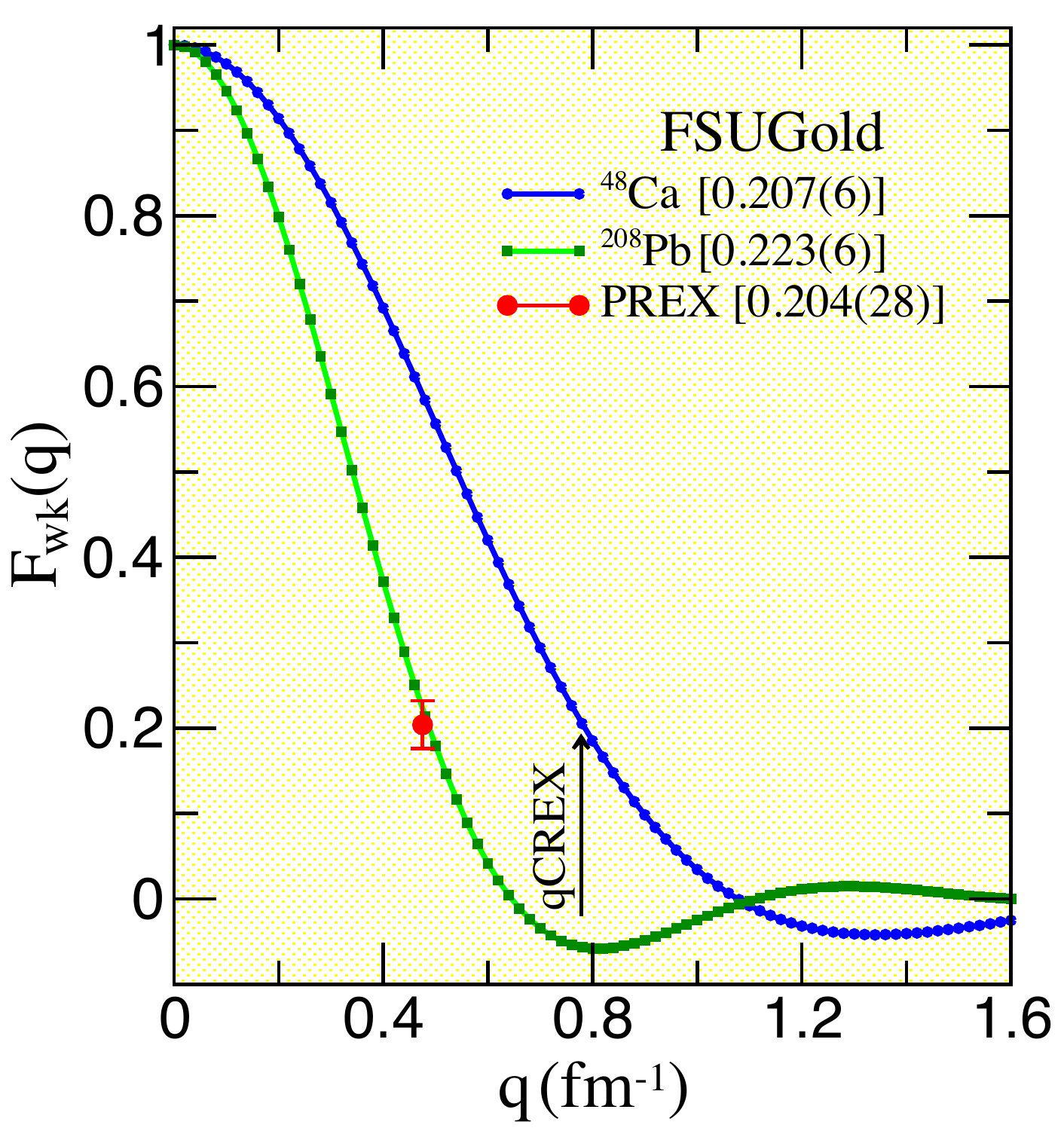}
 \vspace{-0.2cm}
 \caption{(Left panel) Predictions from 48 nuclear density functionals for the neutron-skin thickness 
               of ${}^{48}$Ca and ${}^{208}$Pb\,\cite{Piekarewicz:2012pp}. Constraints from 
               planned experiments have been incorporated into the plot. (Right panel) Weak
               form factors for ${}^{48}$Ca and ${}^{208}$Pb as predicted by the FSUGold
               density functional\,\cite{Reinhard:2013}. The PREX data point is also shown.} 
  \label{Fig1}
\end{figure}

On the left-hand panel of Fig.\,\ref{Fig1} we display predictions from various 
accurately calibrated models for the neutron-skin thickness of both 
${}^{48}$Ca and ${}^{208}$Pb\,\cite{Piekarewicz:2012pp}. Unlike the case 
of other heavy neutron-rich nuclei, 
the correlation between ${}^{48}$Ca and ${}^{208}$Pb is not very tight. This
is because the ratio of surface to volume is significantly larger in ${}^{48}$Ca 
than in ${}^{208}$Pb. This further strengthen the case for measuring two 
neutron-rich (doubly-magic) nuclei on opposite ends of the nuclear chart. 
Also shown in the plot are the projected error bars for PREX-II ($\!\pm0.06$\,fm),
CREX ($\!\pm0.02$\,fm)\,\cite{CREX:2013}, and for a possible  
measurement at the {\sl Mainz Energy-Recovering Superconducting Accelerator}
(MESA) Facility in Mainz ($\!\pm0.03$\,fm)\,\cite{Sfienti:2013}. Finally, we note 
that although most of the models displayed in the figure predict accurately the 
binding energy and charge radii of a variety of nuclei, they are unable to agree 
on whether ${}^{48}$Ca or ${}^{208}$Pb has the larger neutron skin.

As we leave this section it is important to clarify the connection between
the experiment and the theoretical extraction of the neutron radius. 
Although some mild model-dependent assumptions must be made in going from 
the experiment to $R_{n}$\,\cite{Horowitz:2012tj}, such assumptions are ultimately 
unnecessary. This is because the measured parity-violating asymmetry (in Born 
approximation) is proportional to the weak form factor $F_{\rm wk}(Q^{2})$---which is 
obtained directly from the {\sl theoretical} weak-charge density via a Fourier transform. 
Indeed, both the weak-charge radius and $F_{\rm wk}(Q^{2})$ are obtained from 
suitable integrations over the calculated weak-charge density\,\cite{Reinhard:2013}. 
Such an example is provided on the right-hand panel of Fig.\,\ref{Fig1}, where the 
weak form-factor of ${}^{48}$Ca and ${}^{208}$Pb are displayed as predicted by the 
FSUGold density functional\,\cite{ToddRutel:2005zz}. Quantities in parenthesis 
provide the corresponding theoretical and experimental errors. The PREX data 
(at $q\!=\!0.475\,{\rm fm}^{-1}$) and projected location of the CREX data  
(at $q\!=\!0.778\,{\rm fm}^{-1}$) are also shown.

%%%%%%%%%%%%%%%%%%%%%%%%%%%%%%%%%%%%%%%%%%%%%%%%%%%%%%%%%%%%%%
%%%%%%%%%%%%%%%%%%%%%%%%%%%%%%%%%%%%%%%%%%%%%%%%%%%%%%%%%%%%%%

\section{Neutron Stars}

Besides the fundamental role that neutron skins play in elucidating the isovector sector
of density functionals, neutron skins are also critical in constraining the equation of state 
(EOS) of neutron-rich matter, and ultimately the structure, dynamics, and composition 
of neutron stars. In particular, the neutron-rich skin of heavy nuclei (such as ${}^{208}$Pb)
is strongly correlated to a fundamental parameter of the EOS: the slope of the symmetry 
energy at saturation density (a quantity customarily denoted by $L$). Note that the symmetry 
energy may be regarded as the difference in energy between 
pure neutron matter and symmetric matter. Given that symmetric nuclear matter saturates, 
namely, its pressure vanishes at saturation density $\rho_{{}_{0}}$, $L$ is directly related to 
the pressure of pure neutron matter: $P_{{}_{\rm PNM}}(\rho_{{}_{0}})\!=\!\rho_{{}_{0}}L/3$. 
When wiewed in this context, the strong correlation between the neutron skin of ${}^{208}$Pb
and $L$ is easily understood: the larger the value of the symmetry pressure $L$ the stronger
the push against surface tension, and thus the thicker the neutron skin\,\cite{Horowitz:2001ya}.
Such a strong correlation between $r_{\rm skin}^{208}$ and $L$ is nicely illustrated on the
left-hand panel of Fig.\,2, where a large number of density functionals were used to predict
both\,\cite{RocaMaza:2011pm}. The strong correlation of 0.979 suggests that a laboratory 
observable such as $r_{\rm skin}^{208}$ may serve as a proxy for a fundamental property 
of the EOS. Also shown in the plot are anticipated errors for PREX-II and for the corresponding
experiment at the proposed MESA facility\,\cite{Sfienti:2013}. An experiment with such a 
small error bar could significantly constrain the slope of the symmetry energy. 

%%%%%%%%%%%%%%%%%%%%%%%%%%%%%%%%%%%%%%%%%%%%%%%%%%%%%%%%%%%%%%
\begin{figure}[h]
 \includegraphics[height=7cm]{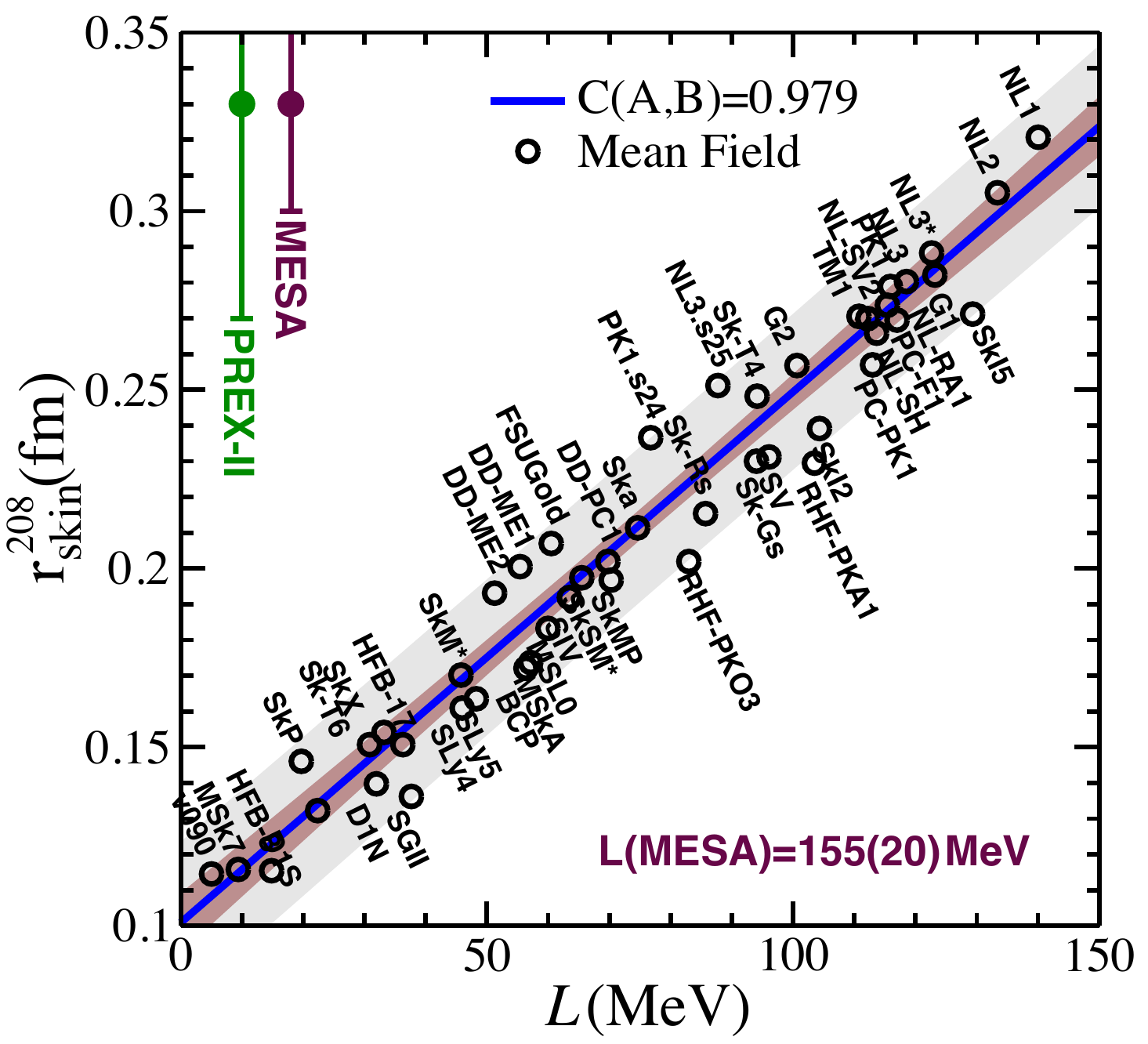}
  \hspace{0.75cm}
 \includegraphics[height=7cm]{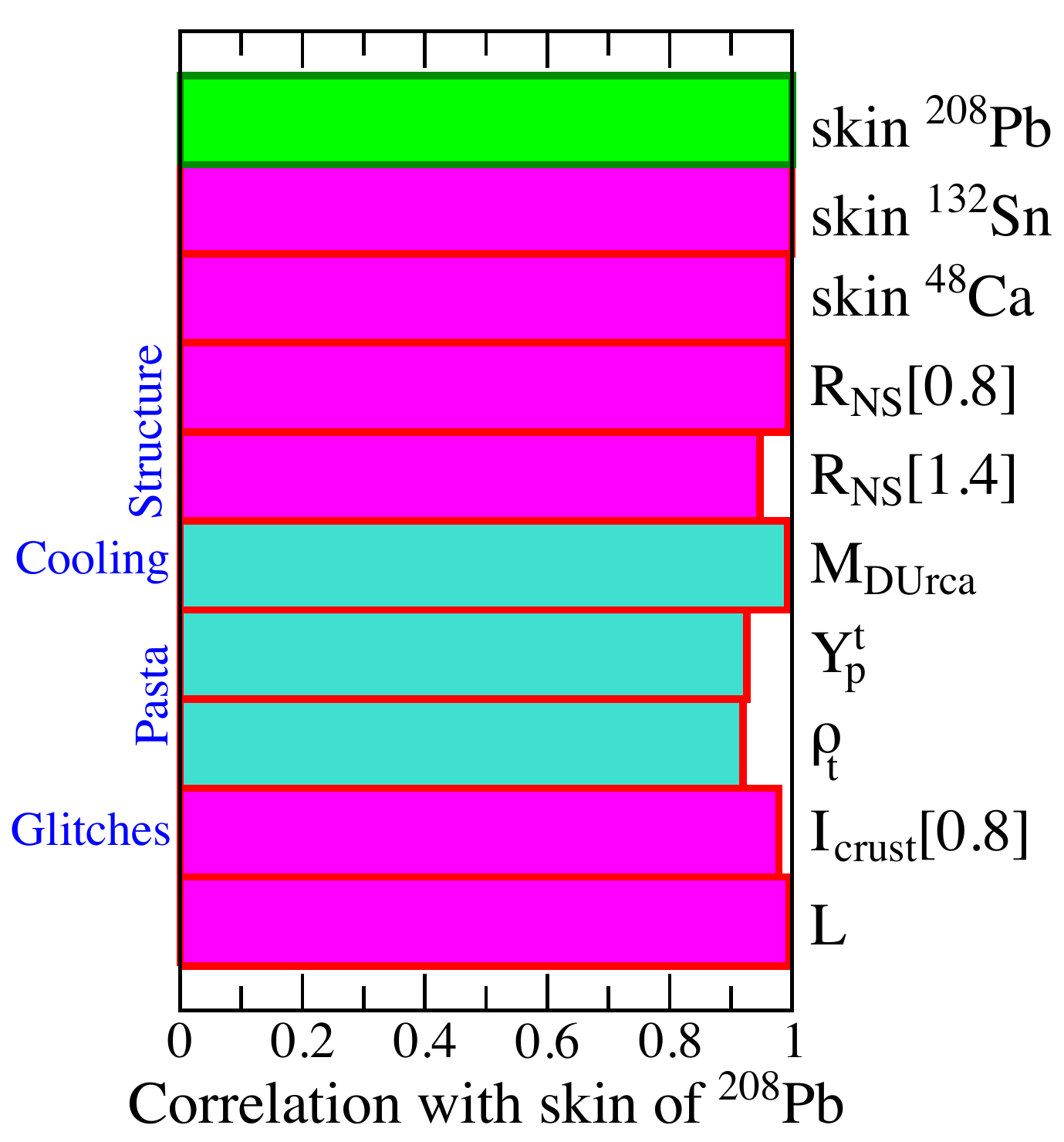}
 \vspace{-0.2cm}
  \caption{(Left panel) Predictions from a large number of nuclear density functionals for the 
  	       neutron-skin thickness of ${}^{208}$Pb and the slope of the symmetry
	       energy $L$\,\cite{RocaMaza:2011pm}. Constraints from planned 
	       experiments have been incorporated into the plot. (Right panel) Correlation
	       coefficients between $r_{\rm skin}^{208}$ and a variety of neutron-star
	       properties as predicted by the FSUGold density functional.} 
 \label{Fig2}
\end{figure}
%%%%%%%%%%%%%%%%%%%%%%%%%%%%%%%%%%%%%%%%%%%%%%%%%%%%%%%%%%%%%%

It is interesting to note that the same pressure that pushes against surface tension 
and creates a neutron-rich skin in nuclei also determines the radius of a neutron star. 
Although a neutron star contains regions of density significantly higher than those 
encounter in a nucleus, it has been shown that stellar radii are controlled by the density 
dependence of the symmetry energy in the immediate vicinity of nuclear-matter saturation
density\,\cite{Lattimer:2006xb}. This provides a powerful {\sl ``data-to-data''} relation: 
The smaller the neutron-skin thickness of ${}^{208}$Pb the smaller the radius of a neutron
star\,\cite{Horowitz:2001ya}.

Besides the stellar radius there are many other neutron-star properties that are sensitive to 
the density dependence of the symmetry energy. In particular, a quantity that is highly 
sensitive to $L$ and that strongly impacts the structure, dynamics, and composition of a
neutron star is the proton fraction $Y_{\rm p}\!=\!Z/A$. Given that $L$ controls the rate of change 
of the symmetry energy near saturation density, models with a soft symmetry energy (namely, 
one that changes slowly with density) tolerate proton fractions at high density significantly lower 
than their stiffer counterparts. Conversely, those same soft models predict larger proton fractions 
at sub-saturation density than their stiffer counterparts. In particular, models with a stiff symmetry
energy predict large proton fractions in the high-density stellar core. Such large values of $Y_{\rm p}$
may be sufficient to explain the enhanced cooling of certain neutron stars via the direct Urca
process\,\cite{Horowitz:2002mb}. However, if the proton fraction in the stellar core is small as
a result of a soft symmetry energy, the observed enhanced cooling may require the presence
of exotic cores\,\cite{Page:2004fy}. The very strong correlation between a large number of stellar
properties and the neutron-skin thickness of ${}^{208}$Pb (or equivalently $L$) is illustrated on 
the right-hand panel of Fig.\,\ref{Fig2}. Such a correlation was investigated with the FSUGold model by 
performing a proper covariance analysis\,\cite{Fattoyev:2012rm}. In addition to the correlation
between $r_{\rm skin}^{208}$ and the neutron-skin thickness of other neutron-rich nuclei, 
Fig.\,\ref{Fig2}. displays correlations between $r_{\rm skin}^{208}$ and radii of neutron stars of different 
masses\,\cite{Carriere:2002bx}, properties at the crust-core boundary\,\cite{Horowitz:2000xj},
and crustal moment of inertia\,\cite{Fattoyev:2010tb}. Note that the threshold direct-Urca mass
$M_{\rm DUrca}$, and the proton fraction $Y_{\rm p}^{\rm t}$ and density $\rho_{{}_{\rm t}}$ at the 
crust-core boundary are strongly {\sl anti-correlated} to $r_{\rm skin}^{208}$. 

%%%%%%%%%%%%%%%%%%%%%%%%%%%%%%%%%%%%%%%%%%%%%%%%%%%%%%%%%%%%%%
\begin{figure}[h]
 \includegraphics[width=7cm]{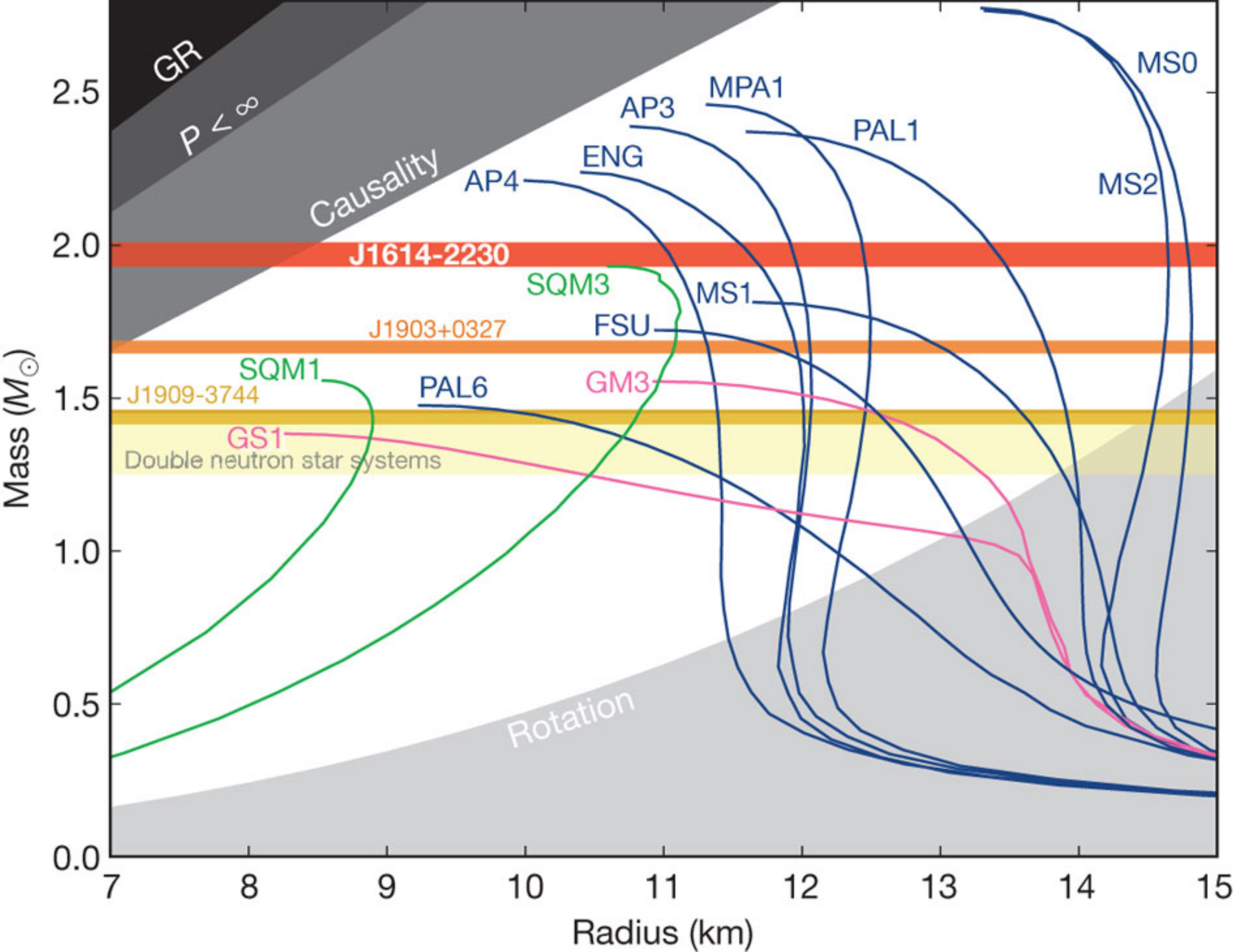}
  \hspace{0.75cm}
 \includegraphics[width=7cm,height=5.5cm]{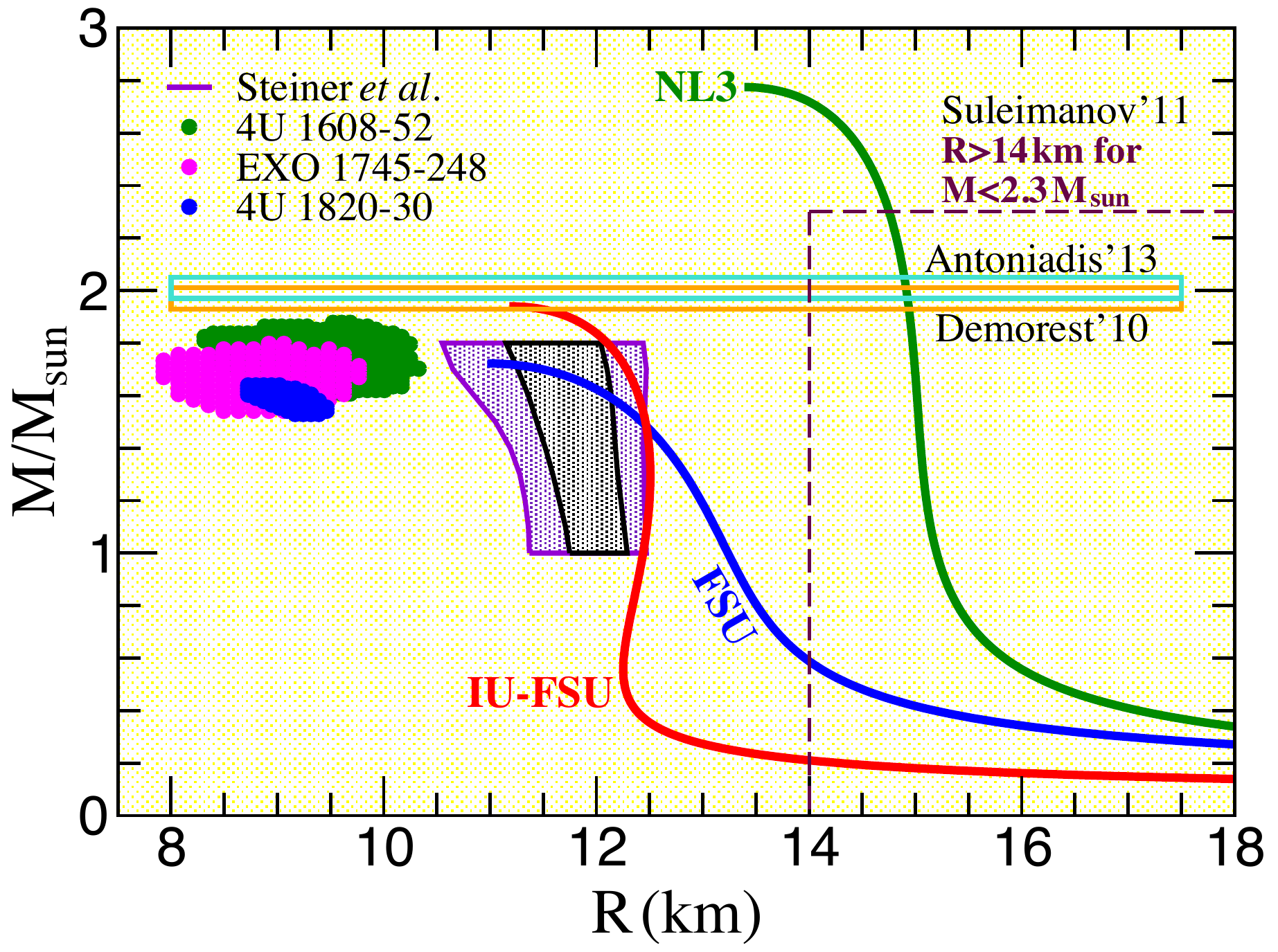}
 \vspace{-0.2cm}
   \caption{(Left panel) Predictions for the mass-vs-radius relation of neutron stars for
                a variety of models using both exotic and non-exotic 
                cores\,\cite{Demorest:2010bx}. The figure illustrates the virtue
                of measuring large mass neutron stars in constraining the equation
                of state. (Right panel) Constraints on stellar radii extracted from
                various analysis of X-ray bursts, as described in the text.}
 \label{Fig3}
\end{figure}
%%%%%%%%%%%%%%%%%%%%%%%%%%%%%%%%%%%%%%%%%%%%%%%%%%%%%%%%%%%%%%

We finish this contribution by addressing briefly the impact of astrophysical observations on the 
behavior of neutron-rich matter. To do so we show on the left-hand panel of Fig.\,\ref{Fig3} 
mass-vs-radius relations as predicted by a variety of models---including those containing strange-quark 
matter. Also displayed on the figure with horizontal bars are well-measured neutron star masses---including 
the large $M\!=\!(1.97\!\pm\!0.04)\,M_{\odot}$ mass reported by Demorest and 
collaborators\,\cite{Demorest:2010bx}. Note that at the time of this writing Antoniadis and collaborators 
reported the existence of the most massive (albeit only marginally) neutron star to date: 
$M\!=\!(2.01\!\pm\!0.04)\,M_{\odot}$\,\cite{Antoniadis:2013}. Clearly, theoretical models that predict 
limiting masses below $2\,M_{\odot}$ require an adjustment of the high-density component of the
EOS. Undoubtedly the search for even more massive neutron stars will continue and this will provide 
invaluable insights into the behavior of high-density matter.

Whereas laboratory experiments are of little value in constraining the EOS of cold, fully catalyzed, 
neutron-rich matter at high densities, we have argued that the measurement of the neutron-rich skin of 
heavy nuclei is critical in constraining neutron-star radii. We now address the complementary 
question: how do neutron-star radii constrain the slope of the symmetry energy $L$, and thus ultimately 
the neutron-skin thickness of ${}^{208}$Pb. Recently, significant advances in X-ray astronomy have 
allowed the simultaneous determination of masses and radii from a systematic study of several X-ray 
bursters~\cite{Ozel:2010fw, Steiner:2010fz}. Results from such studies are displayed on the
right-hand panel of Fig.\,\ref{Fig3}. In particular, the  analysis from Ozel and collaborators
(occupying the $8$-to-$\!10$ km region) suggests very small radii that are difficult to reconcile with the 
predictions from models lacking exotic cores~\cite{Fattoyev:2010rx}. In contrast, a later analysis 
by Steiner, Lattimer, and Brown that addresses systematic uncertainties suggests larger stellar 
radii\,\cite{Steiner:2010fz}. However, even this more conservative estimate has been called
into question\,\cite{Suleimanov:2010th}. Indeed, the authors of Ref.~\cite{Suleimanov:2010th} 
propose a lower limit on the stellar radius of 14\,km for neutron stars with masses below 
2.3\,M$_{\odot}$, thereby suggesting a fairly stiff symmetry energy. Although we are confident
that thermal emissions during X-ray bursts will become a powerful tool for determining stellar
radii, we believe that laboratory experiments on neutron-rich skins provide at present the best
alternative.

%%%%%%%%%%%%%%%%%%%%%%%%%%%%%%%%%%%%%%%%%%%%%%%%%%%%%%%%%%%%%%
\begin{theacknowledgments}
  This work was supported in part by grant DE-FD05-92ER40750 from the Department of Energy.
\end{theacknowledgments}
%%%%%%%%%%%%%%%%%%%%%%%%%%%%%%%%%%%%%%%%%%%%%%%%%%%%%%%%%%%%%%

\bibliographystyle{aipproc}

\end{document}